\documentstyle[12pt,epsf,supcite]{article}
\def\thebibliography#1{\section*{References\markboth
 {References}{References}}\list
 {\arabic{enumi}.}{\settowidth\labelwidth{[#1]}\leftmargin\labelwidth
 \advance\leftmargin\labelsep\itemsep 0pt
 \usecounter{enumi}}
 \def\newblock{\hskip .11em plus .33em minus -.07em}
 \sloppy \sfcode`\.=1000\relax}

\begin{document}

\title{\large\bf SIX-QUARK CONFIGURATIONS IN THE NN \\
SYSTEM CORRELATED WITH EXPERIMENT}
\author{V. S. Gorovoy and I. T. Obukhovsky}
\date{}
\maketitle
\begin{center}Institute of Nuclear Physics \\
Moscow State University, Moscow 119899, Russia
\end{center}
\bigskip
\begin{abstract}
The nucleon-nucleon interaction at short range is analyzed in terms of
six-quark configurations on a base of numerical solutions of modified
RGM equations. It is shown that in low partial waves L=0,1
the system has a two-channel character: the $NN$ channel and the inner
six-quark state ("bag") with specific color-spin structure.
Starting from this analysis it is shown that
polarization observables could be a good tool for investigation of a quark
structure of the deuteron.
\end{abstract}
\section{Introduction}
Now it is still impossible to deduce the nucleon-nucleon interaction,
nucleon-meson form factors and other hadron properties directly
from the QCD and we are forced to use for these purposes
models inspired by QCD (effective colorless field, constituent quarks,
chiral solitons, {\it etc.}). In our opinion the quark model approach
to the intermediate energy nucleon-nucleon dynamics is more
appropriate than the standard meson-exchange diagram technique, which
is inefficient in this area because of a large uncertainty of
meson-nucleon form factors and a basic inadequacy of the meson
perturbation theory. Considering nucleons as three-quark clusters
we can reduce complex problems of nucleon-nucleon and meson-nucleon
dynamics to more simple ones that could be solved with well known
algebraic methods of the nuclear clustering theory~\cite{jphys,echaya,nemetz}~.

On a base of the quark-model approach it was previously
shown~\cite{tamag,ob79,kus91}
that in the $NN$ system at short range (in low partial
waves $L=0,1$) the excited six-quark configurations $s^4p^2\,\,(L=0)$
and $s^3p^3\,\,(L=1)$ play a key role because of their non-trivial
permutation symmetry (Young schemes $[42]_X$ and $[3^2]_X$ correspondingly).
The point is that in these configurations the every color-spin (CS) state
from Clebsch-Gordon series of inner production of color (C) and spin (S)
Young schemes
$$
[f_{CS}]=\cases{[2^3]_C\circ [42]_S=[42]_{CS}+[321]_{CS}+
[2^3]_{CS}+[31^3]_{CS}+[21^4]_{CS}\,, &S=1\,,\cr
[2^3]_C\circ [3^2]_S=[3^2]_{CS}+[41^2]_{CS}+[2^21^2]_{CS}+
[1^6]_{CS}\,, &S=0\cr}\eqno (1)
$$
satisfies the Pauli exclusion principle in all spin-isospin channels
(S,T)=(1,0), (0,1), (0,0), (1,1). So we can construct a
fully antisymmetrised (over quark permutations) nucleon-nucleon wave
function at short range on a base of excited configurations only.
On the contrary the Pauli principle constraints ("Pauli
blocking") play a crucial role for non-excited configurations $s^6\,(L=0)$
and $s^5p\,(L=1)$. It is a consequence of a very simple permutation symmetry
of configurations $s^6$ and $s^5p$ (Young schemes $[6]_X$ and $[51]_X$) at
which most of CS states from Eq.(1) do not satisfy the Pauli exclusion
principle.

The significance of excited six-quark configurations for $NN$ interaction at
short range was confirmed by many authors (see e.g. Ref.~\cite{faes}
with resonating
group method (RGM) calculations at law energy $E_{lab}<0.5\:GeV$ and by our
modified RGM calculations~\cite{kus91} in a large interval of energy
$0<E_{lab}<1.5\:GeV$. However it is still impossible to correlate excited
($s^4p^2$, $s^3p^3$) or non-excited ($s^6$, $s^5p$) configurations with any
observable effects. This is a general problem of "visualization" of
quark degrees of freedom in nuclear phonomena  at intermediate energy.

In this report we should like to show  that in experiments with polarized
deuteron beams there are a possibility to observe effects which is connected
with specific spin structure of excited  six-quark configurations in the
deuteron. The non-excited configurations could be correlated with observed
structures in the energy dependence of spin asymmetry  of NN scattering
cross section  at intermediate energy as it was shown in
works~\cite{lom,kal}.

\goodbreak
\section{Modified RGM approach}

As it was said above in the lowest six-quark configurations $s^6[6]_XL$=0
and $s^5p[51]_XL$=1 most of the CS states from Eq.(1) are forbidden and the
excited configurations $s^4p^2[42]_XL$=0 and $s^3p^3$$[3^2]_XL$=1 play an
important role in the $NN$ interaction at short range. If allowed states
in the configurations $s^6$ and $s^5p$ ({\it e.g.} $s^6[6]_X[2^3]_{CS}
LST$=010 in the $^3S_1$ channel) do not play a crucial role in the $NN$
dynamics at short range we can isolate them in the $NN$ function (in the
first approximation), and the remainder would be a nodal $NN$ wave function.
It could justify a good description of the $NN$ scattering in terms of
Moscow potentials with forbidden states in low partial waves
$L=0,\,1$~\cite{kuku85,dor91} and show that there is a simple origin of the
repulsive core in the $NN$ system independently on the form of $qq$
interaction and just as a consequence of the Pauli exclusion principle.
To verify such picture of $NN$ interaction at short range the sophisticated
numerical RGM calculations were recently made~\cite{kus91}.

The large basis of six-quark configurations was incorporated into
the RGM wave function by using the two-centered shell-model basis,
$$
\mid S^3_+(R)S^3_-(R)[f_X][f_{CS}]LST>\equiv\mid S^3_+S^3_-\{f\}LST>\,,
{\ }{\ }\{f\}\equiv \{[f_X],[f_{CS}]\}\eqno (2)
$$
where $S_{\pm}({\bf R})={\cal N}\exp[-{1\over{2b^2}}
\sum_{i=1}^3({\bf r}_i\mp{{\bf R}\over 2})^2]$.
The trial wave function of our modified RGM is taken in the form of an
integral over the generator coordinate R and a sum over Young schemes ~$f$
$$
\Psi^{RGM}_{NN}(1,2,...6)=
\sum_f\int\limits_0^{\infty}
\chi^L_f(R)\,[{\cal I}^L_{f_X}(R,R)]^{-{1\over 2}}
\mid S^3_+(R)S^3_-(R)\{f\}LST>
R\,dR\,,\eqno (3)
$$
$$
\chi^L_f(R)=x^L_f{\delta(R)\over R}+
U^{NN}_f[\tilde N^L_{f_X}(k,R)+
\cot\delta_L\tilde J^L_{f_X}(k,R)]\,,\eqno (4)
$$
where $\tilde N^L_{f_X}(k,R)$ and
$\tilde J^L_{f_X}(k,R)$ are generator coordinate analogies of
spherical Bessel functions ${\it n}_L(kR)$
and ${\it j}_L(kR)$, which correspond to free asymptotics of $NN$
scattering.
In the Eq.(4) the $U^{NN}_f$ is a unitary matrix of transformation
from Young schemes $\{f\}$ to
quantum numbers of the $NN$ channel. Coefficients $x^L_f$ and phase
shifts $\delta_L$ are unknown values, which must be calculated by the
solution of the RGM equations at fixed energies of the $NN$ system
$E_{NN}={{k^2}/{m_N}}={{k^2}/{3m_q}}$
The $\delta$-functions in the trial functions Eq.(4) incorporate
six-quark shell-model configurations $s^6[6]_X\{f\}$,
$s^4p^2[42]_X\{f\}$, {\it etc.} to the RGM wave function Eq.(3).
It is well known~\cite{har}, that two-centered configurations Eq.(2)
in the limit $R\to 0$ become standard shell-model states with the
same Young schemes $\{f\}$:
$$
\lim_{R\to 0}\,[{\cal I}^L_{f_X}]^{-{1\over 2}}
\mid S^3_+(R)S^3_-(R)\{f\}LST>=
\cases{
\mid s^6[6]_X\{f\}LST>\,,\,\, if\, [f_X]=[6]\,,\cr
\mid s^4p^2[42]_X\{f\}LST>\,,\,\, if\, [f_X]=[42]\cr}\eqno (5)
$$
We reduce the RGM equations to a system of linear algebraic equations
for $x^L_f$ and $\cot\delta_L$
and the solution of the modified RGM equations has a form of a
superposition of the six-quark configurations $s^6$ and $s^4p^2$ and
the asymptotical RGM states in the $NN$ channel
$$
\Psi^{RGM}_{NN}(1,2,...,6)_{L=0}=
x^{L=0}_{[6]}(k)\mid s^6[6]_X[2^3]_{CS}LST=010>
$$
$$
+\sum_{f_{CS}}x^{L=0}_{[42]_X[f_{CS}]}(k)
\mid s^4p^2[42]_X[f_{CS}]LST=010>
$$
$$
+\sqrt{10}A\left\{N(1,2,3)N(4,5,6)(1-e^{-{{r^2}\over{4\gamma^2}}})
[n_L(kr)+\cot\delta_Lj_L(kr)]\right\}_{L=0}\eqno (6)
$$

\section{Analysis of the NN wave function at short range
on a base of microscopical calculations}
{\bf The model of interaction. }
We make use of the quark Hamiltinian
$$
H_q=\sum_{i=1}^6(m_q+{{p^2_i}\over{2m_q}})+
\sum_{i>j=1}^6(V^{OGE}_{ij}+V^{Conf.}_{ij}+V^{II}_{ij}+
V^{Ch.}_{ij})\,,\eqno (7)
$$
where $V^{\alpha}_{ij}$, $\alpha=OGE,...,Ch.$ are effective $qq$
potentials of perturbative QCD interaction
$$
V^{OGE}_{ij}=\alpha_s{{\lambda_i\lambda_j}\over 4}
[{1\over{r_{ij}}}-{\pi\over{m^2_q}}(1+{2\over 3}{\bf\sigma}_i{\bf\sigma_j})
\delta({\bf r}_{ij})-{1\over{4m^2_q}}{1\over{r^3_{ij}}}
S_{ij}({\bf\hat r}_{ij})]\,,\eqno (8)
$$
phenomenological confinement
$$
V^{Conf.}_{ij}=-{{\lambda_i\lambda_j}\over 4}(g_cr_{ij}-V_0),\eqno (9)
$$
and non-perturbative QCD motivated

({\it i}) instanton-induced ($II$) interaction~\cite{dorokh93}
$$
V^{II}_{ij}=-y\rho^2_c{{\pi}\over 3}
({{1-\tau_i\tau_j}\over 4})
[1+{3\over 8}{{\lambda_i\lambda_j}\over 4}
(1+3{\bf\sigma}_i{\bf\sigma}_j)]\delta({\bf r}_{ij})\,,\eqno (9)
$$
where $\rho_c=0.3\,fm$ is the instanton radius and y is an adjustable parameter, and

({\it ii}) effective interaction of quarks with $\pi$ and $\sigma$
mesons inspired by NJL (Nambu-Jona-Lasinio) model of
spontaneous breaking chiral symmetry (SBCS),
$$
H_{Ch.}=g_{\pi qq}F(Q){\bar \psi}(\sigma+
i\gamma_5{\bf\tau\pi})\psi\,\eqno (11)
$$
where form factor F(q) is of the form
$F(Q)=[1+\sum_{k=1}^nC_k{{Q^2+m^2_{\pi}}
\over {Q^2+\Lambda^2_k}}]^{1\over 2}$
that is convenient for coordinate representation of $V^{Ch.}_{ij}$.
As usual $m_{\sigma}=2m_q$, $m_q={1\over 3}m_N$,
$\alpha_{Ch.}\equiv{{g^2_{\pi qq}}\over{4\pi}}
{{m^2_{\pi}}\over{4m^2_q}}$=
$\left({3\over 5}\right)^2{{g^2_{\pi NN}}\over{4\pi}}
{{m^2_{\pi}}\over{4m^2_N}}=0.0276$.
Free parameters of interactions in Eq.(8) -(11), $\alpha_s$, $g_c$,
$y$ and $V_0$, were fitted to the spectrum
of non-strange baryons of positive parity with two variants for
$\pi qq$ form factor F(Q):

({\it a}) a "hard" form factor with momentum cut off at
$Q_c\approx 1.5\,GeV/c$,

({\it b}) a "soft" form factor with
$Q_c\approx{1\over{\rho_c}}\approx 0.6\,GeV/c$ (in this case the form
factor F(Q)
approximates the function ${{m_q(Q)}\over{m_q(0)}}$ where $m_q(Q)$ is
the momentum dependent constituent quark mass from Diakonov's theory of light
quarks in the instanton vacuum~\cite{diak}).
Without any additional fitting of parameters we obtained not so bad
description of the $^3S_1$ phase shift of $NN$ scattering in a large interval
of energy $0<E<1\,\,GeV$ (see fig. 1).

In contrast to the earlier RGM calculations in
which authors considered only low energy $NN$ scattering up to $400\,MeV$
our approach was constructed to be feasible for intermediate energy
applications too (for example the baryon spectrum in table 1 covers the range
at least $0.5-1\,GeV$ in mass).

\medskip
\centerline{{\bf Table 1.} Two variants of quark-quark potentials }
\medskip

\begin{tabular}{cccccccc}
&Parameters of interaction&&&r.m.s.&&Baryon spectrum&\\
\end{tabular}
\smallskip

\begin{tabular}{ccccccccccc}
\hline
&$\alpha_s$&$g_c$&$V_0$&y&b&N&$N^{**}_{{1\over 2}^+}$&
$\tilde N^{**}_{{1\over 2}^+}$&$\Delta$&$\Delta^{**}$\\
Var. &&(${{Mev}\over{fm}}$) & (MeV) && (fm) & \\[5pt]
\hline
({\it a}) & 0.42 & 49.4 & 26.0 & 3. & 0.54 & 939 & 1437 & 1703 & 1230 &
1863 \\
({\it b})& 0.44 & 152.6 & 144.6 & 7. & 0.56 & 939 & 1449 & 1715 & 1230 &
1839 \\[5pt]
\hline
&&&&&Exp.:& 939 & 1440 & 1710 & 1232 & 1600 \\
&&&&&&& $\pm40$ & $\pm30$ & $\pm2$ &  $\pm150$ \\
\end{tabular}

\begin{figure}[h]
\epsfxsize=10cm
\hfill \epsfbox{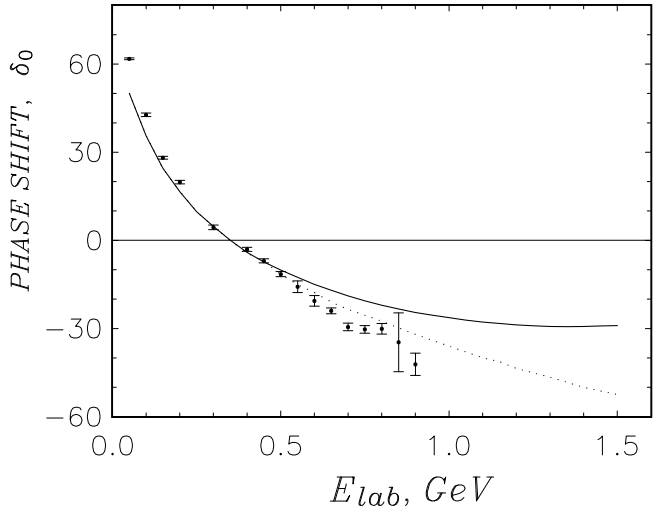} \hfill {\ }
\caption
{$^3S_1$ phase shift of $NN$ scattering. RGM calculations (var. a),
solid; dynamical reconstruction on a base of Moscow
potential~\protect\cite{dor91}, dots; phase analysis from
Ref.~\protect\cite{arn}}
\end{figure}

\noindent
 Of course it would be unreasonable to expect
a quantitative description of $NN$ scattering above $0.5-1\,GeV$ in any
quark microscopical approach. But we can
make some qualitative analysis of behavior of the system at short range
on a base of microscopical calculations and it would help to develop a reasonable phenomenological
potential model, which takes into account quark degrees of freedom.

\medskip
{\bf NN short range behavior. }
The results on the $NN$ wave function in an overlap region are shown
in the fig. 2 (a) and (b).
The $NN$ component of the six-quark wave function can be obtained by
projecting the RGM solution Eq.(6) into the $NN$ channel by using the
fractional parentage technique (f.p.c.)~\cite{jphys,echaya,kus91}
$$
\Phi^{L=0}_{NN}(r)=\sqrt{{6!}\over{3!3!}}<N(1,2,3)N(4,5,6)\mid
\Psi^{RGM}_{NN}(1,2,...,6)>\eqno (12)
$$
The $\Phi_{NN}(r)$ is suppressed at short distances (see fig. 2 (a)) and looks
like a wave function of a standard phenomenological potential model with
the repulsive core.

\begin{figure}[h]
\epsfxsize=10cm
\hfill \epsfbox{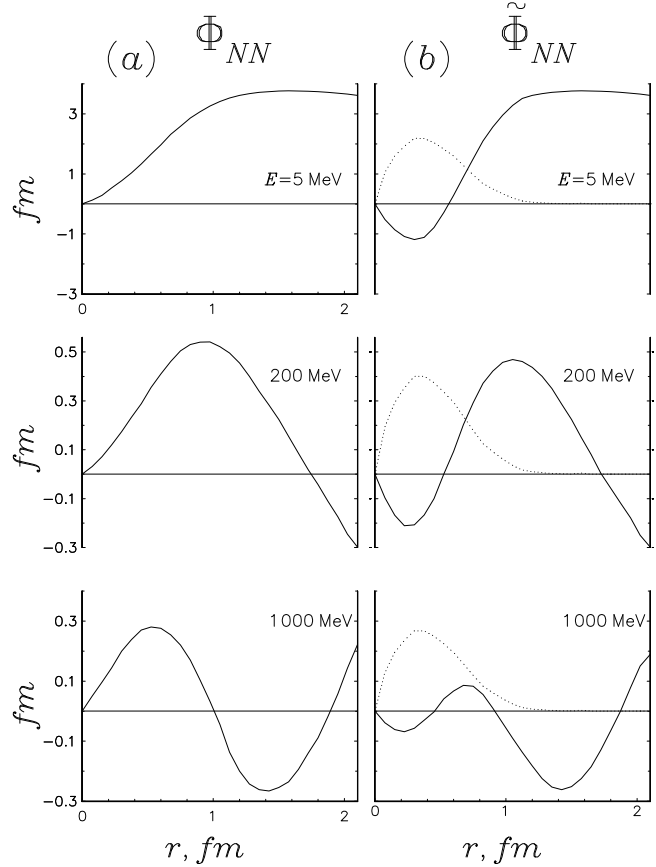} \hfill {\ }
\caption{Results of RGM calculations~\protect\cite{kus91} at energies
$E_{lab}$=5, 200 and 1000 $MeV$. Projections of the
$\Psi^{RGM}_{NN}$ into $NN$ channel: (a) full
function $\Phi^{L=0}_{NN}$, (b) reduced function
$\tilde\Phi^{L=0}_{NN}$ (solid) and $s^6$ configuration (dots). }
\end{figure}

However, if we subtract the contribution of the
configuration $s^6$ (the first term at the r.h.s. of Eq.(6)), we obtain
a nodal wave function
$$
\tilde\Phi^{L=0}_{NN}(r)=\sqrt{{6!}\over{3!3!}}<N(1,2,3)N(4,5,6)\mid
\left[\mid\Psi^{RGM}_{NN}(1,2,...,6)>\right.
$$
$$
\left.-x^{L=0}_{[6]_X}\mid s^6[6]_X[2^3]_{CS}LST=010>\right]\eqno (13)
$$
with a stable position of the node at distances
$r\approx b\approx 0.5$-$0.6\,fm$ in a large interval of energy
$0<E<1\,GeV$ (see fig.2 (b)). The $\tilde\Phi_{NN}(r)$ is orthogonal to
the inner $0S$ state originating from the configuration
$s^6[6]_X$, which is not in essence the $NN$ state: this configuration has approximately
identical projections into any one of possible baryon-baryon channels
of given $S$ and $T$ ($NN$,$\Delta\Delta$, $CC$, {\it etc.}). The $P$-wave
has also a nodal $NN$ structure at short range (after subtraction of a
contribution of the configuration $s^5p$ from the full wave function
$\Psi^{RGM}_{NN, L=1}$).

Then we can identify the function $\tilde\Phi^{L=0}_{NN}(r)$ as "a true
$NN$ state" in the open $NN$ channel and the configuration $s^6$ as
"a six-quark bag state" in the closed channel. From our microscopical quark
calculations it follows (see fig. 1 and 2 (b)) that we can describe
a repulsive core
behavior of the $NN$ phase shifts in $S$ and $P$ waves taking into account
only the open $NN$ channel with its nodal wave function at short range
and using (as the first approximation to $NN$ interaction) in this channel
an attractive $NN$ potential with the forbidden $0S$ and $1P$ states ("Moscow
potential"~\cite{kuku85,dor91}). But to take into
account a specific non-local character of the $NN$ interaction at short
range we have to make use of a coupled closed channel (as a second approximation)
with its symmetrical six-quark wave function. Note that the closed six-quark
channel is orthogonal to the $NN$
state of the open channel. Now on a base of this consideration
there is proposed a new (more adequate than standard OBEP
or Moscow~\cite{kuku85,dor91} potentials) phenomenological
description of the $NN$ interaction at short range~\cite{kuku95}.

\goodbreak

\section{Six-quark structure of the deuteron}
\nobreak
The deuteron $S$-wave function consists also of two orthogonal components:
(i) the inner part, $s^6$ bag (closed channel), (ii) the outer part, an
orthogonal to the $s^6$ bag nodal $NN$ wave function. The deuteron $D$ wave
is another $NN$ component, connected with $s^6$ bag through the $qq$ tensor
forces and with $S$-wave $NN$ component --- through the tensor force of the
pion-exchange potential. Table 2 represent the full solution for the
deuteron $S$ and $D$ waves on the basis of 13 quark configurations.

\medskip
\centerline{{\bf Table 2.} Quark configurations in the deuteron.}
\medskip
\centerline{ $S$ wave:}
\medskip
\begin{tabular}{cccccccc}
\hline
Conf.&$s^6$&$s^4p^2$-$s^52s$&&&$s^4p^2$&\\
&$[6]_X$&$[6]_XL$=0&&&$[42]_XL$=0&\\[5pt]
\hline
$[f_{CS}]$&$[2^3]_{CS}$&$[2^3]_{CS}$&$[42]_{CS}$&$[321]_{CS}$&$[2^3]_{CS}$&
$[31^3]_{CS}$&$[21^4]_{CS}$\\
\hline
$\Psi_{6q}$&0.155&-0.019&-0.115&0.042&0.052&-0.01&0.004\\
\hline
f.p.c.&&&&&&\\
$NN$&$\sqrt{\frac19}$&$\sqrt{\frac1{45}}$&$\sqrt{\frac1{25}}$&
-$\sqrt{\frac{64}{2025}}$&-$\sqrt{\frac1{405}}$&$\sqrt{\frac2{405}}$&0\\
$\Delta\Delta$&-$\sqrt{\frac4{45}}$&-$\sqrt{\frac4{225}}$&0&0&
-$\sqrt{\frac{64}{2025}}$&0&$\sqrt{\frac4{81}}$\\
$N^*N$&0&0&-$\sqrt{\frac1{50}}$&-$\sqrt{\frac2{2025}}$&
$\sqrt{\frac5{162}}$&$\sqrt{\frac1{405}}$&0\\
\hline
\end{tabular}
\medskip

\goodbreak

\centerline{ $D$ wave:}
\nobreak
\medskip
\begin{tabular}{ccccccc}
\hline
Conf.&$s^4p^2$-$s^52s$&&&$s^4p^2$&\\
&$[6]_XL$=2&&&$[42]_XL$=2&\\[5pt]
\hline
$[f_{CS}]$&$[2^3]_{CS}$&$[42]_{CS}$&$[321]_{CS}$&$[2^3]_{CS}$&
$[31^3]_{CS}$&$[21^4]_{CS}$\\
\hline
$\Psi_{6q}$&0.007&0.045&-0.023&-0.011&0.009&-0.002\\
\hline
f.p.c.&&&&&&\\
$NN$&$\sqrt{\frac1{45}}$&$\sqrt{\frac1{25}}$&
-$\sqrt{\frac{64}{2025}}$&-$\sqrt{\frac1{405}}$&$\sqrt{\frac2{405}}$&0\\
$\Delta\Delta$&-$\sqrt{\frac4{225}}$&0&0&
-$\sqrt{\frac{64}{2025}}$&0&$\sqrt{\frac4{81}}$\\
$N^*N$&0&0&0&0&0&0\\
\hline
\end{tabular}
\bigskip

The solution differs from standard $NN$ models of the
deuteron in baryon-baryon (BB) and
partial wave composition and has a non-trivial spin structure. This is
essential for deuteron break-up reactions and we expect that new spin
structure of the deuteron wave function could be
observed in experiments with polarized deuteron beams.

{\bf New features of the deuteron wave function.}

1) {\bf Non-nucleon components.} Calculation of the overlap integral
$$
\Phi_{B_1B_2}(r)=\sqrt{\frac{6!}{3!3!}}<B_1(123)\mid<B_2(456)\mid
\Psi_{6q}(12...6)>\eqno (14)
$$
by means of f.p.c. technique~\cite{jphys,kus91} results
the following spectroscopic factors for $\Delta\Delta$, $N^*N$,...{\it etc.}
components in the deuteron:
$S_{\Delta\Delta}=\int
\mid\Phi_{\Delta\Delta}(r)\mid^2\,d^3r=0.02,\quad
S_{N^*N}=\int
\mid\Phi_{N^*N}(r)\mid^2\,d^3r=0.006,...{\it etc.}$
A momentum distribution of these components $\bar\Phi_{B_1B_2}({\bf k})=
\frac1{(2\pi)^3}\int\Phi_{B_1B_2}({\bf r})e^{i{\bf kr}}\,dr^3$ could be observed in
deuteron break-up reactions as momentum distribution of the
baryon-spectator.

2) {\bf Partial wave structure. P-wave components.}  In the six-quark deuteron the nucleon-spectator can originate
not only from $NN$ component but also from $N^*N$ component in which $N^*$ is
the negative parity resonance $N^*_{\frac12^-}(1535)$ and the nucleon is in
a P-wave state~\cite{kus91,kob94,sit}. In the six-quark deuteron there are
two different P- ($N^*N$) states:

a) $^1P^*_1$ state $v_0(r)$ (with S=0) in the configuration
$s^4p^2[51]_X[2^21^2]_{CS}LST$=$100$; by our results its probability
is very small, $<10^{-5}$, and this configuration has a zero
projection into $NN$ channel;

b) $^3P^*_1$ state $v_1(r)$ (with S=1) in the configuration
$s^4p^2[42]_X[42]_{CS}LST$=$010$ with a probability of $\approx0.5\%$

In the deuteron break-up
reaction the outgoing nucleon at fixed angle $\theta$ "remembers" a partial
wave and spin structure of the deuteron (if the initial deuteron is
polarized).
Fourier transform of the wave functions of $S$- and $D$- waves in the
deuteron ($u(k)$ and
$w(k)$) and $P$-wave functions ($v_0(k)$ and $v_1(k)$) are used in
relativistic impulse approximation (RIA)~\cite{kob83} to define the
differential cross section (see details in Ref.~\cite{kob83,kob94,sit})
$$E_q\frac{d^3\sigma}{dq^3}=\frac12C_d\sigma_{NA}\frac1{1-x}
\sqrt{\frac{m_N^2}{4x(1-x)}}
\left[u^2(k)+w^2(k)+C_{off}\left(v^2_1(k)+v^2_0(k)\right)\right]\eqno (15)
$$
the $T_{20}$ analyzing power
$$
T_{20}(0^\circ)=\frac1{\sqrt{2}}
\frac{2\sqrt{2}u(k)w(k)-w^2(k)+C_{off}\left(v^2_1(k)-2v^2_0(k)\right)}
{u^2(k)+w^2(k)+C_{off}\left(v^2_1(k)+v^2_0(k)\right)}\eqno (16)
$$
and the coefficient of polarization transferred
$$
\kappa_0(0^\circ)=\frac{u^2(k)-w^2(k)-
C_{off}\frac3{\sqrt{2}}v_1(k)v_0(k)}
{u^2(k)+w^2(k)+C_{off}\left(v^2_1(k)+v^2_0(k)\right)}\eqno (17)
$$
in the inclusive reaction of the deuteron break-up on nuclei at intermediate
energy. Here $C_{off}$ is a unknown coefficient $0\le C_{off}\le 1$ that
takes into account off-shell effects in the $N^*N$ channel. We choose
an extreme variant of P-wave structure of the deuteron with
$v_0(k)=v_1(k)$ (microscopic calculations give us $v_0\approx 0$, but these
calculations are not adequate to the problem in the case of configuration
$s^4p^2[51]_XL$=1, which has a minimal coupling with $NN$ channel)

\begin{figure}[h]
\epsfxsize=8cm
\hfill \epsfbox{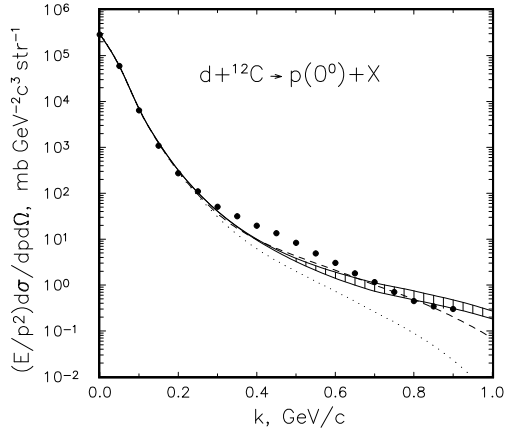} \hfill {\ }
\caption{Inclusive cross section of the deuteron break-up at the Dubna
energy $8.9\,GeV$~\protect\cite{able}. Calculations: six-quarks RGM
without $P$-wave
contribution (lower solid) and with $P$~wave (upper solid), $v_0=v_1$; Paris
potential (dashed); Bonn potential (dots). Stripes
show the region in which $C_{off}$ in Eq. (16) varies from 0 to 1. }

\end{figure}

\newpage

In the figures 3, 4 and 5 there is shown the observables of Eq. (15)-(17),
calculated with
our deuteron wave functions $u$, $w$, $v_0$ and $v_1$ (solid curves).

\begin{figure}[h]
\epsfxsize=8cm
\hfill \epsfbox{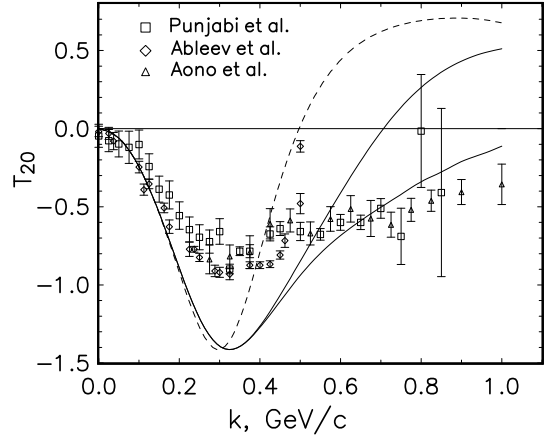} \hfill {\ }
\caption{The $T_{20}$ data~\protect\cite{punj,ableev,aono}. Calculations:
six-quarks RGM without $P$-wave contribution (upper solid) and with $P$~wave
(lower solid), $v_0=v_1$; Paris potential (dashed).}
\end{figure}

\begin{figure}[h]
\epsfxsize=8cm
\hfill \epsfbox{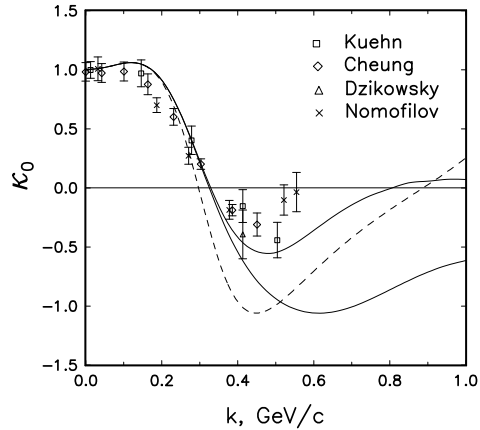} \hfill {\ }
\caption{Polarization transferred data~\protect\cite{kuehn,cheung,dzik,nomof}.
Calculations: six-quarks RGM without $P$-wave contribution (lower solid)
and with $P$~wave
(upper solid), $v_0=v_1$; Paris potential (dashed). }
\end{figure}

Note that in the RIA a momentum ${\bf q} =\{q_{\|},{\bf q}_{\bot}\}$
of the proton-spectator in the deuteron rest frame is correlated
with light-front (l.f.) variables $x$, ${\bf k}_{\bot}=q_{\bot}$
(or l.f. momentum ${\bf k}=\{k_{\|},\,k_{\bot}\}$)
through the equation~\cite{kob83}
$$
x=\frac{\sqrt{m^2_N+{\bf q}^2}+q_{\|}}{2m_N}=
\frac{\sqrt{m^2_N+{\bf k}^2}+k_{\|}}
{2\sqrt{m^2_N+{\bf k}^2}}\eqno (18)
$$

\section{Conclusion and outlook}
\nobreak
The $NN$ interaction at short range (in $S$ and $P$ waves) is significantly
non-local but this non-locality can be effectively described with the
orthogonality condition to the most symmetrical (in X-space) configurations,
$s^6[6]_XL=0$ and $s^5p[51]_XL=1$. These configurations are not
in essence $NN$ states and they have a specific color-spin structure. So
some non-trivial effects could be observed in polarization experiments on
the deuteron ( or in NN scattering) at intermediate energies. We have above
demonstrated that polarization data are sensitive to the inner quark
structure of the NN system.

This approach would be useful for example in
description of "structures" observed in transverse
or longitudinal spin asymmetry of the $NN$ cross-section at intermediate
energies $\Delta\sigma_T=\sigma(\downarrow\uparrow)-
\sigma(\uparrow\uparrow)$, $\Delta\sigma_L=\sigma(^{\gets}_{\to})-
\sigma(^{\to}_{\to})$  and could be considered as a development of
the well known models~\cite{lom,kal}.

One can suppose that the quark Hamiltonian $(16)$ gives an adequate
description of the closed six-quark channel and the interaction that
couples it to the open $NN$ channel, but we expect that
parameters of $qq$ interaction in the six-quark system are not exactly
identical with one of the three-quark system and an
investigation of this problem on
a base of a quantitative model of $NN$ interaction would be very interesting.

\bigskip
\goodbreak
\medskip\noindent
\section{Acknowledgement}
\medskip
\nobreak
\noindent
The authors are very thankful to Prof. A.P.Kobushkin for his critical
comments and help in correction of our error in the Eq. (17).
The author wishes to express his gratitude for fruitful discussions and
suggestions to his colleagues in Moscow University Professors
V. G. Neudatchin and  V.I.Kukulin and Dr. Yu. M. Tchuvil'sky.
This work was supported by the Russian Foundation for Basic Research
(grant 93-02-03375).


\medskip

\begin{thebibliography}{99}

\bibitem{jphys}
I. T. Obukhovsky, Yu. F. Smirnov, and Tchuvil'sky,
{\it J. Phys.} {\bf A15} (1982) 7.
\bibitem{echaya}
V. G. Neudatchin, I. T. Obukhovsky and Yu. F. Smirnov,
{\it EChAYa} {\bf 15} (1984) 1165.
\bibitem{nemetz}
O. F. Nemets, V. G. Neudatchin, A. T. Rudchik, Yu. F. Smirnov
and Yu. M. Tchuvilsky, {\it Nucleon Clustering in Light Nuclei
and Multinucleon Transfer Reactions},
(in russian, Naukova Dumka, Kiev, 1988).
\bibitem{tamag}
V.G. Neudatchin, Yu.F. Smirnov and R. Tamagaki,
{\it Progr.Theor.Phys.} {\bf 58} (1977) 1072.
\bibitem{ob79}
I. T. Obukhovsky {\it et al.},
{\it Phys. Lett.} {\bf B88} (1979) 231.
\bibitem{kus91}
A. M. Kusainov, V. G. Neudatchin, and I. T. Obukhovsky,
{\it Phys. Rev.} {\bf C44} (1991) 2343.
\bibitem{faes}
A. Faessler {\it et al.},
{\it Phys. Lett.} {\bf B90 } (1982) 41.
\bibitem{lom}
E. Lomon, {\it Physique} {\bf 58} (1990) {\it Supple.
Colloque C6}, 363.
\bibitem{kal}
Yu. S. Kalashnikova, I. M. Narodetsky, and Yu. A. Simonov,
{\it S. J. Nucl. Phys.} {\bf 46} (1987) 689.
\bibitem{kuku85}
V. I. Kukulin {\it et al},
{\it Phys. Lett.} {\bf B153} (1985) 7.
\bibitem{dor91}
Yu. L. Dorodnykh, V. G. Neudatchin {\it et al.},
{\it Phys. Rev.} {\bf C43 } (1991) 2499.
\bibitem{har}
M. Harvey, {\it Nucl. Phys.} {\bf A352 } (1981) 301.
\bibitem{dorokh93}
A. Dorokhov and S. Kochelev,
{\it Phys. Lett.} {\bf B204 } (1993) 167.
\bibitem{diak}
D. I. Diakonov, V. Yu. Petrov, and P. V. Pobylitsa,
{\it Nucl. Phys.} {\bf B306} (1988) 809.
\bibitem{arn}
R. A. Arndt, J. S. Hyslop III, and L. D. Roper,
{\it Phys. Rev.} {\bf D35} (1983) 128.
\bibitem{kuku95}
V. I. Kukulin, (1995). {\it Private communication}, not to be published.
\bibitem{kob94}
A. P. Kobushkin, in {\it 14th Int. IUPAP Conf. on Few Body Problems in
Physics}, (Williamsburg, Virginia, U.S.A.,
1994), p.~99.
\bibitem{kob83}
A. P. Kobushkin and V. P. Shelest, {\it EChAYa} {\bf14} (1983) 1146
\bibitem{sit}
M. P. Rekalo and I. M. Sitnik, {\it Phys. Lett.} {\bf B354} (1995) 434;
I.M.Sitnik, V.P.Ladygin, M.P.Rekalo,
{\it Preprint} E1-94-23; E1-94-154, (JINR, Dubna, 1994).
\bibitem{able}
V. G. Ableev {\it et al., Pis'ma ZhETF} {\bf 37} (1983) 196;
{\bf 45} (1987) 467; {\it Nucl. Phys.} {\bf A393} (1983) 491; {\bf A411}
(1984) 541(E).
\bibitem{punj}
V. Punjabi {\it et al., Phys. Rev.} {\bf C39} (1990) 608.
\bibitem{ableev}
V. G. Ableev {\it et al., Pis'ma ZhETF} {\bf 47} (1988) 558;
{\it JINR Rapid Comm.} {\bf4[43]-90} (1990) 5.
\bibitem{aono}
T. Aono {\it et al., Preprint} DPNU-94-36, (Nagoya, 1994).
\bibitem{kuehn}
B. K\"uhn {\it et al., Phys. Lett.} {\bf B334} (1994) 298.
\bibitem{cheung}
N. E. Cheung {\it et al., Phys. Lett.} {\bf B284} (1992) 210.
\bibitem{dzik}
T. Dzikovsky {it et al.}, in {\it Proc. Int. Workshop
DUB\-NA-DEU\-TE\-RON'91}, E2-92-25,
(JINR, Dubna, 1992), p.~181.
\bibitem{nomof}
A. A. Nomofilov {\it et al., Phys. Lett.} {\bf B325} (1994) 327.


\end{thebibliography}
\end{document}